# Molecular Beam Study of the CO Adsorption on a Regular Array of PdAu Clusters on Alumina

Georges Sitja* and Claude R. Henry*
Aix Marseille Université / CNRS, CINAM, campus de Luminy, case 913
F-13288 Marseille cedex 09, France

*Corresponding Authors: henry@cinam.univ-mrs.fr, sitja@cinam.univ-mrs.fr

## Abstract

The adsorption kinetics of CO on PdAu bimetallic clusters, containing 140 ± 12 atoms and a composition varying between 0% and 55% of Pd atoms, is investigated by a pulsed molecular beam method (MBRS). The clusters are grown on a nanostructured ultrathin film of alumina on $Ni_3Al$ (111) playing the role of a template which gives a hexagonal array of bimetallic clusters having a sharp size distribution and a uniform composition. The surface concentration calculated, assuming segregation of gold to the surface, varies between 0 and 90% of Au atoms on the surface. From the adsorption-desorption kinetics of CO, the lifetime of CO is measured at various temperatures. At low coverage, plotting the CO lifetime in an Arrhenius diagram one obtains the adsorption energy of CO. When the surface concentration of Au increases, the adsorption energy of CO on the PdAu clusters decreases. This evolution of the adsorption energy is discussed, from previous studies, in term of ligand and ensemble effects. We find that the ensemble effect plays a dominant role in the observed decrease of the adsorption energy of CO.

## 1.Introduction

Bimetallic catalysts become more and more important in catalysis because often they present better selectivity, higher stability and sometimes higher activity than the pure metal counterparts. Another reason for the interest of bimetallic catalysts is to decrease the content of rare and expensive precious metals like Pt, Rh or Pd. In particular gold combined with Pd has been extensively studied in the recent years[1,2]. For basic studies, supported model catalysts constituted by metal nanoparticles grown under UHV on oxide surfaces are used for the possibility to characterize them with great details by surface science techniques[3-5]. The group of Hajo Freund has prepared CoPd[6-7], PdAg[8,9] and PdFe[10] bimetallic model catalysts by depositing the two metals on an ultrathin alumina film on NiAl (110). Due to the low thickness of the oxide, the bimetallic nanoparticles could be characterized in-situ by STM. The same group has also prepared PdAu nanoparticles on $CeO_2$, $Fe_3O_4$ and MgO thin films grown respectively on Ru(0001), Pt(111) and Ag(100)[11]. Peter Varga and co-workers have also prepared PdCo clusters on an ultrathin film of alumina on NiAl(110)[12]. Wayne Goodman group has prepared model catalysts of AgAu and PdAu on bulk $TiO_2(110)$[13,14]. The same group has also synthesized AuPd and NiAu model catalysts supported on $SiO_2$ thin films grown on Mo bulk surfaces[15,16]. From these studies of bimetallic model catalysts it turns out that the chemical composition is not uniform among the particles[17]. In the case of simultaneous deposition of the two metals, if the interaction of adatoms with the substrate is different for the two metals (it is generally the case), the nucleation rate of the two metals is different leading to large differences in particle composition. The sequential deposition of the metals can partially answer to the previous problem by a proper choice of the metal which is deposited first. However, in both cases inhomogeneities in composition due to non-uniform spatial distribution of the particles, still remain. To avoid these problems of inhomogeneous composition one can grow sequentially the bimetallic particles on a surface which presents a regular array of nucleation centers. This templated growth has been recently used to form regular arrays of mono-metallic clusters with a narrow size distribution[18]. The templates are ultrathin alumina film on $Ni_3Al$ (111)[19], graphene monolayer on metals[20] and h-BN monolayer on metals[21].

Bimetallic clusters templated growth has been first realized by sequential growth of PdAu clusters on alumina/$Ni_3Al$(111)[22,23] and later PtRu clusters on graphene/Ru(0001)[24] and NiAu clusters on h-BN/Rh(111)[25]. Only the PdAu/alumina/$Ni_3Al$(111) system has been studied in details. From STM and GISAXS studies[26,27], the organization of the PdAu cluster-lattice is stable for clusters from few to 400 atoms. The lattices are stable in temperature up to 600K[26] and under oxygen atmosphere ($10^{-6}$ mbar)[28] and under CO atmosphere[29].

In this paper we study the adsorption of CO on PdAu clusters having the same size and variable chemical compositions. The adsorption energy is measured by a molecular beam method (MBRS : Molecular Beam Relaxation Spectrometry) as it was previously realized for Pd clusters lattices[29].

## 2. Experimental

The alumina films are prepared by oxidation of a clean $Ni_3Al(111)$ surface at 1000K during an exposition of 40L of oxygen under $5 \times 10^{-8}$ mbar followed by an annealing under UHV at 1050 K during 7 min[30]. The films have been characterized by LEED, AFM and STM at RT[26,30]. The surface of the ultrathin film (0.5 nm thick) presents a superstructure forming a hexagonal array with a unit mesh of 4.1 nm. This superstructure is used as a template to grow the bimetallic clusters. Pd was deposited first and then Au at 320K with Knudsen cells calibrated in-situ by a quartz micro-balance, then the quantity of each metal contained in the clusters is accurately known. It has been shown previously that during the first exposure to the Pd beam a hexagonal array of Pd clusters is formed and during the second exposure the Au atoms are selectively deposited on the previously grown Pd clusters, then an array of bimetallic clusters with a chosen composition is obtained[22]. The density of clusters is $6.5 \times 10^{12}$ cm$^{-2}$ as verified by STM observations. It has been previously shown that the number of atoms inside the clusters (n) is known with an accuracy of $\sqrt{n}$[29]. In this study we used PdAu clusters with a mean size of 140±12 atoms containing between 55% and 100% of Pd. For the PdAu particles the accuracy in the chemical composition varies between 3% and 4%.

To study CO adsorption on the PdAu clusters we used the same method as previously for CO adsorption on pure Pd clusters[29]. Briefly, a collimated supersonic beam of pure CO ($2 \times 10^{13}$ molecules cm$^{-2}$ s$^{-1}$ at the sample level) is directed towards the sample and the desorbed CO flux is measured by a differentially pumped quadrupole mass spectrometer[29]. The beam is modulated by a mechanical shutter. At high temperature adsorption is fully reversible then the adsorption - desorption pulses (see Figure 2) can be averaged (typically on 60 individual pulses). Before the first measurement the sample is heated at 550 K in order to equilibrate the PdAu particles and then the temperature is gradually decreased for the adsorption-desorption measurements. From the analysis of the kinetics of the desorption part of the pulses we directly measure the life time of the CO molecule at a given temperature. Then the variations of the life time of the CO molecules as a function of the sample temperature are plotted in an Arrhenius diagram. We keep the temperature sufficiently high to have a low equilibrium coverage of adsorbed CO to avoid a repulsion between adsorbed molecules that would decrease the adsorption energy. This is checked by non-departure from a straight line of the Arrhenius plot and also by the fact that the life time of adsorbed CO is the same in the two half-periods of the pulse (when the CO beam is on we have adsorption and desorption and when the beam is off we have a pure desorption). The slope of the Arrhenius straight line gives the desorption energy of CO. At the difference with TPD, by MBRS we measure a single adsorption energy which corresponds to the most stable site.

## 3. Results

A typical STM image of an array of PdAu clusters is displayed on Figure 1. The organization of the clusters in a hexagonal lattice is clearly visible. Figure 2 displays an example of CO adsorption- desorption pulse of CO from Pd clusters. The CO signal can be decomposed in two parts: a fast variation at the opening or closing of the CO beam ($S_{fast}$) which corresponds to the reflection and desorption from a physisorbed state on the bare alumina surface and a slow variation ($S_{slow}$) due to the adsorption or desorption from the Pd clusters. The separation of the two components is easy because the fast component is a straight line while the slow component varies exponentially[29]. It is noticeable that the amplitude of the slow variation is not a constant, but it increases up to a saturation value when temperature decreases. This increase of the adsorption rate is due to the reverse spillover of CO on the alumina support[18]. The $S_{slow}$ signal varies exponentially with time and the time constant is the life time (τ) of the CO molecules on the metal clusters. The adsorption energy is obtained from the variation of the life time of CO molecules on the metal clusters when the temperature changes. The adsorption energy is plotted as a function of their chemical composition in Figure 3. It decreases from 80 kJ/mole for pure Pd to almost zero for a 50% Pd – 50% Au volume composition.

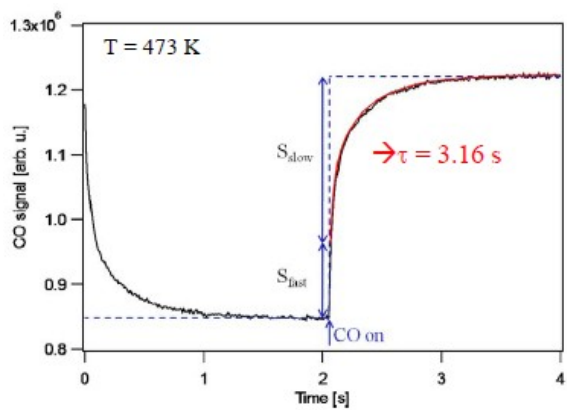

*Figure 1: Example of adsorption-desorption CO signal from an array of Pd clusters on Al2O3/Ni3Al (111).*

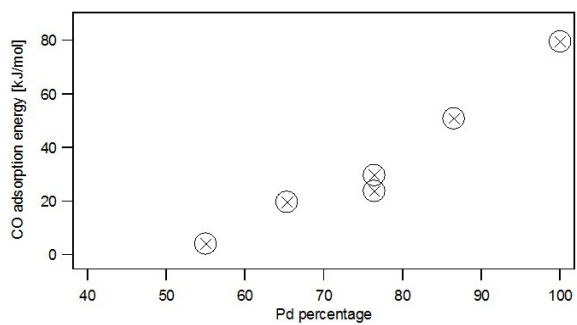

*Figure 2: Adsorption energy of CO, at low coverage, as a function of the chemical composition of the PdAu clusters.*

## 4. Discussion
### 4.1. Surface composition

In order to analyze the variations of the adsorption energy observed on Figure 3 we have to know the concentration of Au and Pd on the surface of the clusters. From thermodynamics, Au is expected to be on the surface because the surface energy of Au is much less than that of Pd[31,32]. This prediction is confirmed theoretically by Monte Carlo simulation[33] and ab-initio

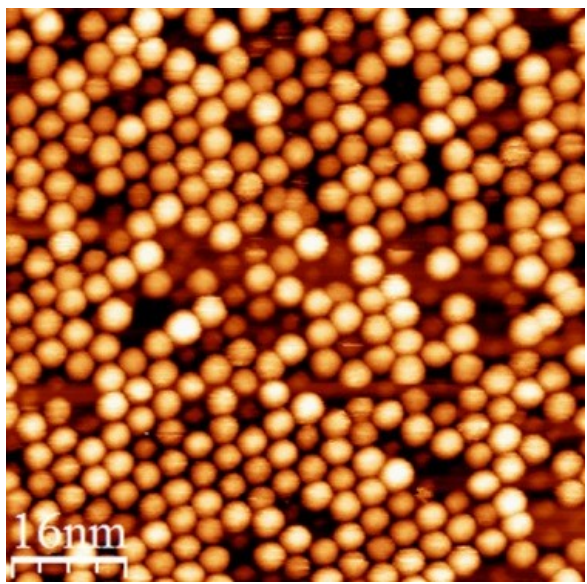

*Figure 3: STM image on an array of Pd50Au50 clusters on Al2O3/Ni3Al (111). Ubias = 1.5V, It = 19pA. The apparent size of the clusters is enlarged by convolution with the STM tip.*

calculation[34]. From the experimental side on PdAu polycrystals[35,36], PdAu single crystals[37-39] and PdAu thin films[40] it has been shown by Auger electron spectroscopy (AES) and ion surface scattering (ISS) that Au segregates to the surface. On surfaces alloys, obtained by depositing Pd (or Au) on Au (or Pd) single crystal surfaces, ISS showed also preferential segregation of gold on the surface[41,42]. Much less studies have been published on the surface composition of supported PdAu particles. For PdAu nanoparticles supported on silica[15] and on an ultrathin alumina film on NiAl (110)[43] it has been shown by ISS that Au segregates on the surface of the nanoparticles whatever Pd or Au was deposited first. The same effect is also predicted theoretically for free PdAu clusters, with gold decorating first the edges[44-46]. In our experiments we deposited first Pd and secondly Au at RT, are we sure that in these conditions the segregation equilibrium is reached ? In fact it has been shown that segregation of gold starts already at RT[47] and is completed at 500 K[41]. Before the CO adsorption experiments the PdAu particles were annealed at 550 K, then we are confident that the segregation equilibrium is reached. Thus before the CO adsorption experiments all the deposited gold atoms are on the surface of the particles because the largest quantity of deposited gold correspond to one

ML assuming hemispherical PdAu particles with a diameter of 1.9nm. However, after CO adsorption do Au atoms stay on the cluster surface ? Indeed, theoretical calculations show that the surface composition depends on the sample temperature and on the CO pressure; at high CO pressure Pd starts to segregate on the surface because the bonding of CO on Pd atoms is much stronger than on Au atoms[48-51]. Pd segregation is predicted to start at $10^{-2}$ mbar[49]. From IR experiments Goodman group deduced that on AuPd (100) segregation of Pd to the surface starts at $1 \times 10^{-5}$ mbar of CO at T>200 K but it becomes important at $10^{-3}$ mbar[52]. On PdAu nanoparticles supported on alumina at 10 mbar of CO Pd segregation was observed at RT by DRIFTS[53]. On $Pd_{70}Au_{30}$ (111) surface it was shown by NAP-XPS that no Pd segregation induced by CO adsorption occurs between $10^{-6}$ and $10^{-2}$ mbar of CO at RT[54]. Another NAP-XPS experiment on AuPd (110) shows Pd enrichment on the surface at a pressure of CO larger than $10^{-2}$ mbar at RT[55]. Thus in our experimental conditions: $P_{CO} = 6.8 \times 10^{-8}$ mbar (equivalent pressure in the beam at the sample surface) and 300< T <600 K Pd segregation induced by CO is not expected. On the contrary we expect that by increasing the proportion of Au deposited on the Pd clusters the surface coverage of gold increases from zero (pure Pd) to one. In table 1 we report the number of gold and of palladium atoms per cluster and the surface concentration in Pd, for the different samples prepared in this study. For the calculation of the surface coverage we take a hemispherical shape for the particles as deduced from previous STM[26] and GISAXS measurements[27,29]. For the size of the cluster we used (D= 1.9 nm) half of the atoms (70) are on the surface. As the maximum number of Au atoms is 63 all the atoms are on the surface of the clusters.

### 4.2. CO adsorption energy

Now we will discuss on the variation of the CO adsorption energy as a function of the chemical composition of the PdAu nanoparticles. The role of alloying two metals on the catalytic properties has been analyzed by two effects: a ligand effect and an ensemble effect[56]. The first effect is related to the modification of the electronic properties of one atom from one kind by the presence of surrounding atoms of the second kind. The second effect, more geometrical, is related to the size of the reaction site which can be constituted by different types of atoms. However the separation of these two effects is not so simple. For example, if we change the size of the ensemble we change the electronic properties of the atoms constituting the ensemble. Liu and Norskov have rationalized the understanding of these two effects by considering the adsorption of carbon monoxide, oxygen and nitrogen from ab initio calculations[57]. The ligand effect corresponds to a continuous shift of the d-band center which is responsible for an almost linear increase of the adsorption energy. The ensemble effect corresponds to the size of the adsorption site (atop site, bridge site or hollow site). In the case of mixed adsorption sites the adsorption energy can be approximated by averaging on the different constituents[57]. In the case of CO on PdAu one can notice that the variation of the adsorption energy for the ligand effect, for the same type of adsorption site, is up to 0.3 eV (29kJ/mol) while it is much larger, 1.5 eV (144kJ/mol), for the ensemble effect (i.e. between the three types of adsorption sites)[57]. The variation of the adsorption energy of CO we observe experimentally is 80kJ/mol (see Figure 3) is too large to be explained by a pure ligand effect.

| Sample | 1 | 2 | 3/4 | 5 | 6 |
|---|---|---|---|---|---|
| $N_{Pd}$ | 140 | 120 | 106 | 91 | 77 |
| $N_{Au}$ | 0 | 20 | 34 | 49 | 63 |
| Volume %Pd | 100 | 86 | 76 | 65 | 55 |
| Surface %Pd | 100 | 71 | 51 | 30 | 10 |

*Table 1: Volume and surface composition of the PdAu nanoparticles grown on $Al_2O_3/Ni_3Al$ (111). The number of Pd atoms ($N_{Pd}$) and gold atoms ($N_{Au}$) in the clusters are also indicated.*

Recent scanning tunneling spectroscopy (STS) experiments on Pd/Au(111) showed that the difference of electronic structure between the Au and Pd atoms is weak and subtle differences can be measured[58]. A very small charge transfer from the Pd atoms toward neighboring Au atoms is observed[58] which is in agreement with earlier XPS measurements during alloy formation in the system Au/Pd(111)[59], although we have to be very cautious in interpreting charge transfer from XPS chemical shifts which are due to the combination of initial and final state effects. The small charge transfer from Pd to Au atoms creates a decrease of the bonding of CO on isolated Pd sites and an increase of the bonding of CO on the neighboring Au atom. These can be attributed to the ligand effect.

Infrared spectroscopy is an efficient technique to probe the geometry of the adsorption site. For CO adsorption on Au/Pd(111)[59] and Pd/Au(111)[60], after a proper annealing to form an equilibrated surface alloy without dissolving the deposited atoms (Au, Pd) into the bulk (Pd or Au single crystals), two peaks are observed corresponding to CO adsorbed on Pd on atop site (2070-2090 cm$^{-1}$) and on bridge site (1910-1980 cm$^{-1}$). The same adsorption sites are also observed on AuPd (111)[61], Au/Pd (100)[62] and AuPd (110)[55] single crystal bimetallic systems. CO adsorbed in hollow site (1810-1850 cm$^{-1}$) is observed only for Au/Pd(111) after annealing at T ≥ 900 K for which Au is dissolved in bulk Pd[59]. Often FTIR spectra are obtained at low temperature (<300K) for which saturation coverage of CO is obtained and no CO in the hollow sites can be observed. However between 350 and 450 K CO in the hollow sites was observed for pure Pd (111) while not on AuPd (111) film in the same conditions[61]. On AuPd particles supported on silica powder at 300K atop and bridge adsorption sites are observed and a weak shoulder corresponding to hollow sites appears for Pd rich bimetallic particles[63]. In the case of AuPd/CeO$_2$(111), AuPd/Fe$_3$O$_4$ and AuPd/MgO (100) model catalysts during CO adsorption at 100K only atop and bridge sites are observed[11]. Even after heating the AuPd/Fe$_3$O$_4$ sample to 400 K after exposure to CO, no CO adsorbed in hollow sites was observed contrary to the case of pure Pd particles[11]. CO adsorption in hollow site is observed only at 450 K or higher temperature on AuPd/Al$_2$O$_3$/Mo(100) model catalysts corresponding to a volume Pd concentration of 83%[64]. However in the last case it is possible that pure Pd particles were also present together with PdAu particles.

The adsorption energy of CO has been determined by TPD (thermal programmed desorption). However, for this technique the adsorption is made at low temperature (typically 100 K) and during the TPD the gas pressure is not present and the CO coverage varies continuously as a function of the temperature, then some site reconversion can occur like change from bridge to hollow site. Thus during TPD one can reveal the different adsorption sites but the coverage is not constant like in infrared spectroscopy. Nevertheless it is a suitable technique to identify the different adsorption sites. On the systems Au/Pd(111)[59] and Pd/Au (111)[65] three desorption peaks at 300/310 K, 350/380 and 445/500 K appear which are identified to atop, bridge and hollow sites, respectively. The corresponding estimated adsorption energies are 76-86, 99 and 117 kJ/mol, respectively. However, for the Au/Pd(111) system[59], the high temperature adsorption peak was not observed after an annealing at 680K. It appears only after annealing at a temperature higher than 800K for which Au is probably dissolved in the Pd bulk. For the Pd/Au (111) system[65] the Pd was deposited at 125 K and CO adsorption was made at the same temperature without annealing. In that case the Pd layer was not equilibrated with underneath gold, an annealing at 600 K is necessary to have gold segregation inside the Pd layer[60]. On AuPd nanoparticles supported on oxide surfaces[11] three TPD peaks appear at 305, 360 and 430 K which are attributed to atop, bridge and hollow adsorption sites[11]. The high temperature peak appears only for a concentration of Pd in the particles of at least 66%[11]. It was noticed that the high energy peak on pure Pd particles was at 450 K (instead of 430K); this observation was explained by a ligand effect. In fact it is difficult to interpret more in detail these experiments on model catalysts because neither the exact composition nor the particle size are known.

The adsorption energy of CO on PdAu surfaces or particles has been also investigated by DFT calculations[49,51,53,57,66-68]. The adsorption energies calculated by DFT are often larger than the experimental adsorption energy and vary from one calculation to another one, mainly because different exchange-correlation functionals are used. For this reason we give also in Table 2 the relative values Ead* referred to a hollow site on Pd (111). From Table 2 we see two main features. First the adsorption energy in a hollow site on a PdAu surface is always lower than on the same site on a pure Pd surface: this fact is a clear evidence of a ligand effect. Secondly by increasing the size of the isolated ensemble of Pd atoms from i=1 to i=3 the adsorption energy of CO increases: this fact is the signature of an ensemble effect. We can notice that the amplitude of the ligand effect is less (about two times smaller) than those of the ensemble effect as it was already discussed before. In the case of AuPd clusters consisting of 38 and 55 atoms, the same hierarchy between the different site coordination is observed but the adsorption energies and the relative values are generally higher than for the extended surfaces[53,58].

| i | 1 | 2 | 3 | Reference |
|---|---|---|---|---|
| site | atop | bridge | hollow | |
| $E_{ad}$ Pd(111) | 96 | 134.5 | 150 | 57 |
| $E_{ad}$ Pd/Au(111) | 81.7 | 91.3 | 105.7 | 57 |
| $E_{ad}*$ | 0.54 | 0.60 | 0.70 | |
| $E_{ad}$ Pd(111) | 94.2 | 133.6 | 150 | 66 |
| $E_{ad}$ Pd/Au(111) | 67.3 | 86.5 | 106.7 | 66 |
| $E_{ad}*$ | 0.45 | 0.58 | 0.71 | |
| $E_{ad}$ Pd(111) | 133.6 | 178 | 198 | 67 |
| $E_{ad}$ PdAu(111) | 118 | 148 | 185 | 67 |
| $E_{ad}*$ | 0.59 | 0.74 | 0.93 | |
| $E_{ad}$ Pd(111) | 113.4 | 137.4 | 165.3 | 49 |
| $E_{ad}$ PdAu(111) | 87.4 | 117.2 | 151 | 49 |
| $E_{ad}*$ | 0.53 | 0.71 | 0.91 | |
| $E_{ad}$ Pd$_{55}$ | | | 212 | 68 |
| $E_{ad}$ (PdAu)$_{55}$ | 134 | 168 | 202 | 68 |
| $E_{ad}*$ | 0.63 | 0.79 | 0.95 | |
| $E_{ad}$ Pd$_{38}$ | 128 | 191 | 204 | 53 |
| $E_{ad}$ (PdAu)$_{38}$ | 96 | 149 | | 53 |
| $E_{ad}*$ | 0.47 | 0.73 | | |

*Table 2: Adsorption energy of CO on Pd ensembles, containing i atoms, calculated by DFT for Pd and PdAu systems. The adsorption energies are in kJ/mol. Ead* is the adsorption energy on a Pd-Au system divided by the adsorption energy in hollow site on pure Pd calculated in the same work. The last four raws correspond to 55 atoms icosahedra and 38 atoms truncated octahedra.*

On Table 3 we give the values of the adsorption energies measured for the different composition by the MBRS method. The clusters are assumed to be hemispheric and for the considered size half of the atoms are on the surface. The total number of atoms is 140 and the fraction of Au atoms in the clusters varies from 0 to 45%. The surface composition is also given in Table 3. For pure Pd clusters the adsorption energy of CO (80kJ/Mol) is much smaller than those measured by the same method on a Pd (111) surface which is 133 kJ/mol[70]. In fact the low value for the 2 nm Pd clusters supported on Al$_2$O$_3$/Ni$_3$Al (111) is a pure size effect which has been investigated in detail on the same system by MBRS[29] and on Pd/Fe$_3$O$_4$/Pt (111) by microcalorimetry[71]. This decrease has several possible origins. First it could be due to the contraction lattice of the Pd particles when the size decreases. This contraction is due to the surface stress which is proportional to the reciprocal radius[72]. The effect on the chemisorption energy can be also understood by a shift of the *d*-band center toward the Fermi level when cluster size increases due to the elongation of Pd-Pd distance[73]. Second, the Pd lattice parameter can also decrease by a positive misfit between the Pd lattice and the support lattice as previously suggested[71]. In the case of Pd clusters supported on NiAl (110) a contraction of the Pd lattice, when the particle size decreases, has been observed by HRTEM[74] and we can assume that the same contraction exist also on $Ni_3Al(111)$. Another effect which has been suggested by the group of Hajo Freund[71] is the decrease of the polarizability of the small particles leading to a decrease the van der Waals attraction between CO molecule and the metal clusters. In our experiment we are always at low coverage then we can safely tell that the Pd is adsorbed on a hollow site on (111) facets that are dominant on these particles growing in (111) epitaxy[27].

| $E_{ad}$(kJ/mol) | 80 | 51.6 | 24.3/30.2 | 20 | 4.15 |
|---|---|---|---|---|---|
| Surface %Pd | 100 | 71 | 51 | 30 | 10 |
| $E_{ad}*$ | 1 | 0.64 | 0.30/0.38 | 0.25 | 0.05 |
| site | hollow | bridge | atop | atop | atop/Au edge |

*Table 3: Experimental adsorption energy of CO on the Pd-Au particles as a function of the surface concentration of Pd and site attribution. $E_{ad}*$ is the adsorption energy divided by the adsorption energy on pure Pd particles.*

In the case of PdAu particles, when the concentration of Pd on the surface of the clusters decreases, the CO adsorption energy continuously drops. For a Pd surface concentration of 71% the CO adsorption energy decreases by 36%. This large value cannot be attributed to a pure ligand effect which would result from DFT calculations (see Table 2) to a maximum decrease between 7 to 30 % (see Table 2) for adsorption in hollow site on Pd trimers. Therefore we attribute the decrease of the CO adsorption energy to an ensemble effect: CO is adsorbed on bridge position on an isolated Pd dimer. In such a case calculations predict a decrease between 26 to 42 % (see Table 2) which is in a rather good agreement with the observed value of 36 %. Furthermore, for a Pd surface concentration of 70% a majority of bridge sites is observed by FTIR on Au/Pd(111)[59]. At a Pd surface concentration of 51% we observed experimentally a further decrease of the CO adsorption energy which corresponds to a reduction of 62 to 70 % relatively to pure Pd. We attribute this large value again to an ensemble effect: CO would be adsorbed atop on an isolated Pd atom. The observed relative decrease is larger than those predicted by theoretical calculation (41 to 55 %) but it is in agreement with experimental measurements on PdAu/Mo(110)[61] and on Au/Pd(111)[59] which show that below a Pd surface concentration of 60%, only atop CO is observed. The fact that surface Pd is only present as isolated atoms has been explained by a small repulsion between Pd adatoms[69]. For a Pd surface concentration of 30% a further decrease of the adsorption energy is observed. This decrease is moderate relatively to the previous Pd concentration and could be attributed to a ligand effect. That means that CO would be adsorbed atop on a Pd atom but the environment would be different. For example in the previous case the Pd atoms could have Pd atoms as second neighbors contrary to the present case where all the second neighbors would be Au atoms. For the last measurement which corresponds to a Pd surface concentration of 10% the adsorption energy of CO is very low and it is tentatively attributed to CO adsorbed atop on an Au atom probably on an edge sites. In fact at room temperature CO does not adsorb on a gold surface but it could be adsorbed on a low coordinated site on a gold cluster[75].

## 5. Conclusion

By using the templated growth method we have prepared regular arrays of PdAu clusters, with a sharp size distribution and a well-defined chemical composition, on alumina ultrathin films on $Ni_3Al$ (111). This is a clear advantage against classical preparation methods of bimetallic supported model catalysts. We have chosen a constant mean cluster size of 140 atoms (1.9 nm in diameter) for which half of the atoms are on the surface. In the PdAu clusters Au tends to be on the surface, then the surface concentration of gold can be determined from the volume composition of the clusters. The gold surface concentration was increased from 0% (pure Pd) to 90%. The adsorption energy of CO at low coverage has been measured by a pulsed molecular beam isothermal method (MBRS). The CO equivalent pressure in the beam was too low to expect Pd segregation on the surface of the clusters induced by CO adsorption.

The adsorption energy decreases when the Au surface coverage increases. From the compilation of experimental studies on extended surfaces of PdAu and DFT calculations the present results have been analyzed on the base of ligand and ensemble effects. We find that the ensemble effect is dominant. On pure Pd clusters CO is adsorbed in hollow sites. When the surface concentration of gold increases, CO is first adsorbed in bridge sites and then on atop sites on ensembles constituted by 2 and 1 Pd atoms, respectively. At an Au surface concentration of 90%, a weak adsorption energy is measured which is attributed to CO adsorbed linearly on gold low coordinated atoms.

## Acnowledgements

The authors acknowledge CNRS for supporting this work.

## Reffrences